\documentclass[preprint,prd,aps,nofootinbib]{revtex4}

\usepackage{graphicx}

\begin{document}

\title {\large Phenomenology of quintessino dark matter \\ --- Production
of NLSP particles}

\author{ Xiao-Jun Bi }
\affiliation{ Institute of High Energy Physics, Chinese Academy of
Sciences, P.O. Box 918-4, Beijing 100039, People's Republic of China}

\author{ Jian-Xiong Wang }
\affiliation{ Institute of High Energy Physics, Chinese Academy of
Sciences, P.O. Box 918-4, Beijing 100039, People's Republic of China}

\author{ Chao Zhang }
\affiliation{ Institute of High Energy Physics, Chinese Academy of
Sciences, P.O. Box 918-4, Beijing 100039, People's Republic of China}

\author{ Xinmin Zhang }
\affiliation{ Institute of High Energy Physics, Chinese Academy of
Sciences, P.O. Box 918-4, Beijing 100039, People's Republic of China}

\date{\today}

\begin{abstract}

In the model of quintessino as dark matter particle,
the dark matter and dark energy are unified in one 
superfield, where the
dynamics of the Quintessence drives the Universe acceleration and its 
superpartner, quintessino, makes up
the dark matter of the Universe. This scenario predicts the existence of 
long lived  $\tilde{\tau}$ as the
next lightest supersymmetric particle.
In this paper we study the possibility of detecting $\tilde{\tau}$
produced by the high energy cosmic neutrinos
interacting with the earth matter. By a detailed calculation
we find that the event rate is one to several hundred per year at a 
detector with effective area of $1 km^2$. The study in this paper can be 
also applied for 
models of gravitino or axino dark matter particles.

\end{abstract}

\maketitle

\section {introduction}

Recent astronomical observations strongly support a concordance model 
of cosmology, where 
the bulk of the content of the Universe is comprised of an unknown
dark sector, of which about 23\% is the pressureless dark matter
and about 73\% is dark energy with negative pressure 
that drives the present acceleration of the Universe.
In the literature there are a lot of interesting and compelling 
models of particle physics which
provide candidates for dark matter and dark energy. However, it is always 
much more desirable that a single model explains both for the dark sector. 
Recently, two of us (X.B. and X.Z.) with 
M. Li\cite{quintessino} have 
considered a class of Quintessence models, originally designed for the 
understanding of the current acceleration of the Universe and showed that
after supersymmetrization the fermionic superpartner of the Quintessence, 
the quintessino, serves as a good candidate for the dark matter particle.
Very much like the quarks and leptons in the same group representation in the 
grand unified theories, the components of the dark matter and the dark 
energy belong to one superfield in this model.
In this scenario, the quintessino is the lightest supersymmetric 
particle (LSP) and the next lightest supersymmetric particle (NLSP)
can be either neutralino or slepton, typically the stau. During the
evolution of the universe,
the NLSP freezes out of the thermal background, then decays 
into quintessino which makes up the present dark matter.

Generally the quintessino dark matter particles interact with the ordinary 
matter very weakly, so the traditional techniques for the direct or 
indirect detection of the neutralino dark matter 
would not be useful for detecting quintessino dark matter. 
There are, however, several silent features of the quintessino
dark matter model which make it distinguishable. 
Firstly, we should note that being one superfield the quintessino
has the same interaction with matter as the quintessence and thus be
severely constrained.
As we know that the primary motivation to introduce the quintessence
field is to drive the acceleration of the Universe, so we have to
preserve the properties of quintessence after introducing interactions
between quintessence and the matter.
In order to keep the quintessence potential flat
and avoid the long range force induced by quintessence we impose a global
shift symmetry for its interactions\cite{carroll}, i.e., 
interactions which are invariant under
$Q\to Q+\text{constant}$.
One possible interaction is  proposed in Ref. \cite{carroll}
$\mathcal{L}_{Q\gamma\gamma}=\frac{c}{M_{pl}}QF_{\mu\nu}\tilde{F}^{\mu\nu}$, 
which can be taken to test the quintessence model by observing
the rotation of the
polarization plane of the light from distant sources. Another
interesting form is $\mathcal{L}_{Qff}=\frac{c}{M_{pl}}\partial_\mu QJ^\mu$
with $J^\mu$ the baryon or baryon minus lepton current\cite{lim}
which gives rise to a new mechanism of the baryogenesis.

Secondly, the BBN observation indicates that the light element $^7$Li may be
underabundant compared with the theoretical estimate\cite{olive}.
Decaying particles after BBN might provide one way to solve the 
problem\cite{olive}. 
As shown explicitly in Ref. \cite{quintessino} (and argued for 
the gravitino dark matter models in Ref. \cite{gravitino}) 
the electromagnetic
energy associated with the non-thermal production of quintessino dark 
matter particles can play the role to reconcile the observations and 
the theory.

Thirdly, the quintessino dark matter in this model is produced 
nonthermally. The property  of the quintessino dark matter is 
characterized by the comoving free streaming scale.
Depending on the time when the NLSP decay and the initial
energy of the quintessino when it is produced,
it can be either cold or warm  dark matter.
In the latter case it helps solve the problems of the 
cold dark matter on subgalactic scales\cite{warm}.

Fourthly, this scenario for dark matter predicts the existence of long-lived
NLSP with life time $10^5-10^8$ sec.
For the stau NLSP it can be produced and collected on 
colliders\cite{collider} and the properties of quintessino 
can be studied by examining the stau decay. The scenario with gravitino LSP and
stau NLSP is studied in the literature\cite{collider}. 
For the case of quintessino LSP the stau will have different decay modes.

In the present paper we are not attempting to study all the features
of quintessino dark matter mentioned above, instead we will 
focus on the study of
the NLSP stau production and detection.
We leave other phenomenological studies for future publications\cite{late}. 

The cosmological stau will not survive until today, 
however it can be produced in the extremely high
energy astrophysical processes.  Experimentally
these stau can then be observed by the large cosmic 
particle detectors on the
earth, such as the L3C\cite{l3c}, SuperKamiokande\cite{sk}, 
or several proposed neutrino telescopes\cite{tele}.
Since the stau are much heavier than other stable or long lived
charged particles, such as the electron or the muon,
they are expected to leave distinct trajectory in the detectors. 

There are different sources of the stau fluxes.
However, since the stau has life time about one year
we expect the stau flux from distant objects, such as AGNs, should
be quite low. Therefore, most stau signal will come from the earth.
One stau flux comes from the high energy cosmic neutrino collision 
with the earth matter and the other one comes
from the collision of the high energy
cosmic proton and nuclei with the atmosphere.
In this paper we will present a detailed calculation of the 
stau flux produced in the simpler case by the cosmic neutrino collision
with the earth matter and leave the discussions on
 the atmospheric flux
of stau in another work\cite{late}.
In a recent paper by Albuquerque, Burdman and Chacko 
the production of the NLSP stau 
by the high energy neutrino flux has been proposed as a possibility 
of probing for
the supersymmetry breaking scale using the neutrino telescope\cite{albu}.

The stau flux thus generated depends on three ingredients: the flux of 
the primary cosmic ray of the incident
neutrinos, the cross sections of the neutrino and nucleon and the 
range of the charged stau on the earth. In this paper we will show that
there will be one to several hundred of stau produced per year at a 
detector with effective area of one ${km}^2$. 
The paper is organized as follows:
in Sec. II, we study how quintessino dark matter is produced by the
stau NLSP decay.
In Sec. III, we calculate the neutrino-nucleon interacting cross sections.
In Sec. IV we discuss the energy distribution of $\tilde{\tau}$ and its 
range. In Sec. V, we describe our calculations and
present the results. Finally, Sec. VI is our conclusion. 

\section{Quintessino production from NLSP stau decay}

In Ref. \cite{quintessino} we focus our study of quintessino 
dark matter on the case of neutralino being NLSP. 
Therefore we will first study the viability of 
stau NLSP to produce the quintessino dark matter in this section.

The most general form of the interaction between the quintessence
and the leptons that obeys the shift symmetry is given by\cite{quintessino}
\begin{equation}
\mathcal{L}_{Qff}=\frac{ 1 }{\Lambda}\partial_\mu Q
( c_{ij}^R \bar{f_i}_R\gamma^\mu f_{j R}
+ c_{ij}^L \bar{f_i}_L\gamma^\mu f_{j L}) \ ,
\end{equation}
with $\Lambda$ being the cutoff scale and $c_{ij}$ being coupling
constant in the effective theory. The astrophysical and laboratory
experiments put a lower bound on $\Lambda$ of about $10^{10} GeV$
\cite{quintessino}. Here we take a simple form of the interaction
between quintessence and the lepton $\tau$
\begin{equation}
\mathcal{L}_{Qff}=\frac{ c }{\Lambda}\partial_\mu Q
 \bar{\tau}\gamma^\mu \gamma^5 \tau \ .
\end{equation}
After supersymmetrization and using the on-shell condition of $\tau$
we get the relevant coupling for our discussion
\begin{equation}
\label{qff}
\mathcal {L}_{\tilde{Q}\tau\tilde{\tau}} =\frac{2cm_\tau}{\Lambda}
(\bar{\tilde{Q}} P_L \tau \tilde{\tau}_R^*
+\bar{\tilde{Q}} P_R \tau \tilde{\tau}_L^* )+ h.c.\ ,
\end{equation}
where $\tilde{\tau}_L$ and $\tilde{\tau}_R$ 
are the left- and right-handed stau respectively.

The quintessino is produced by the stau decay, $\tilde{\tau}
\to \tilde{Q}+\tau$.
From Eq. (\ref{qff}), the time scale for the stau decay is given by
\begin{equation}
\tau(\tilde{\tau}\to \tilde{Q} \tau)\approx
\frac{10^6\text{sec}}{c^2} \left(\frac{\Lambda}{\Lambda_{GUT}}
\right)^2 \left(\frac{1 TeV}{m_{\tilde{\tau}}}\right) (1-x^2_Q)^{-2}\ ,
\end{equation}
where $x_Q=m^2_{\tilde{Q}}/m^2_{\tilde{\tau}}$ and $\Lambda_{GUT}
\approx 2\times 10^{16} GeV$ being the scale for grand unification.
Taking $10^{10} GeV\lesssim\Lambda\lesssim 10^{13} GeV$,
the stau decays before BBN. In this case there will be no constraints
on the process from BBN.

For higher scale $\Lambda$, the stau decays after BBN. The electromagnetic
energy released in the stau decay may destroy the successful prediction
on the light elements abundance of BBN\cite{bbn} 
and the Planckian spectrum
of  the cosmic microwave background (CMB)\cite{cmb}.
Therefore we have to consider the constraint from BBN and CMB. 
The electromagnetic energy released in the stau decay can be written as
\begin{equation}
\xi_{EM}=\epsilon_{EM}N_{\tilde{\tau}}
=0.75\times 10^{-9} GeV \frac{1-x_Q^2}{x_Q}\ ,
\end{equation}
where $\epsilon_{EM}$ is the initial electromagnetic energy
released in each stau decay and
$N_{\tilde{\tau}}=n_{\tilde{\tau}}/n_\gamma^{BG}=
3.0\times 10^{-12}
\left[ \frac{TeV}{m_{\tilde{Q}}} \right]
\left[ \frac{\Omega_{DM}}{0.23} \right]$ is the number
density of stau
normalized to the number density of the background photons\cite{gravitino}.
As the argument given in Ref. \cite{gravitino}, we take 
$\epsilon_{EM}=\frac{1}{2}E_\tau$ for simplicity, where 
$E_\tau$ is the energy of $\tau$ in the decay products.

\begin{figure}
\includegraphics[scale=0.65]{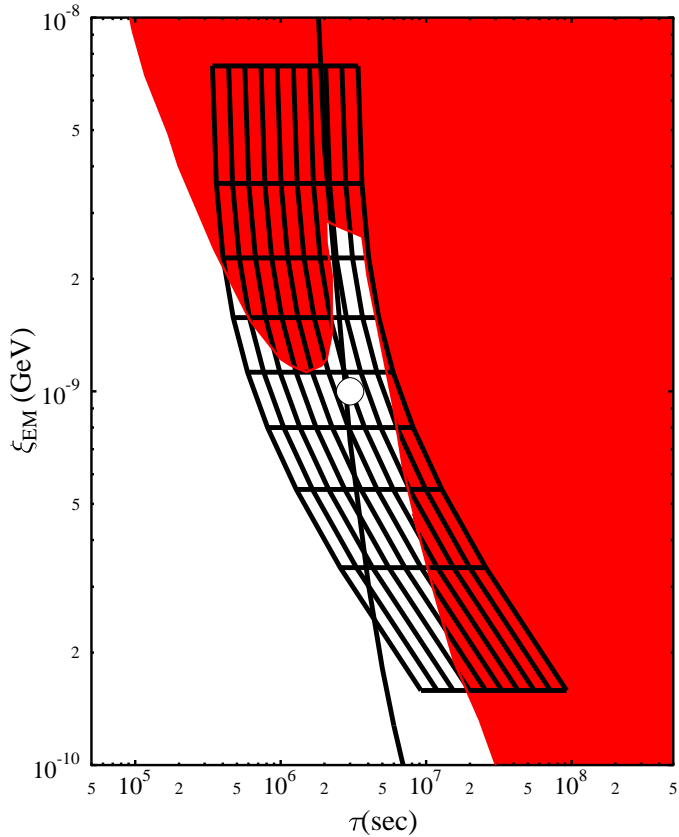}
\caption{\label{para}
Lifetime $\tau$ and energy release $\xi_{EM}$ in stau decay
for $m_{\tilde{\tau}}$ from $300 GeV$ to $ 3 TeV$ (from right
to left) and $x_Q$ from $0.1$ to $ 0.9$ (from top to bottom).
We take $c=1$ and $\Lambda=2\times 10^{16} GeV$.
The contour of the chemical potential for photon
distribution function is $\mu=9\times 10^{-5}$,
with the region to the right of it being excluded by CMB\cite{pdg}.
The shaded regions are excluded by BBN data\cite{bbn}.
The circle represents the best fit region with
$(\tau, \xi_{EM})= (3\times 10^6 sec, 10^{-9} GeV)$\cite{bbn}.
}
\end{figure}

In Fig. \ref{para}, we plot the decay time and energy release
for $m_{\tilde{\tau}}$ from $300 GeV$ to $3 TeV$ and $x_Q$
from $0.1$ to $0.9$. The shaded region is excluded by BBN.
The best fit point $(\tau,\xi_{EM})=(3\times 10^6\text{sec},10^{-9} GeV)$
is covered in the parameter space,
which can suppress the level of $^7$Li and make it
to be consistent with the observation
\cite{bbn}.

We will not further consider the constraint from BBN on the stau 
three-body hadronic decay\cite{hadr} in this work. Compared with the recent 
analysis given by Feng \textit{et. al}\cite{feng}, we find there is
viable parameter space around the best fit point even after imposing
the hadronic constraint in our model. 
Therefore, we get the conclusion that the quintessino dark matter
produced by stau decay is a viable dark matter model, which is consistent
with the present experimental constraints.

\section{supersymmetric neutrino-nucleon cross section}

Probing for the long lived stau provides an indirect support for our 
dark matter model\footnote{
In the case of the gravitino\cite{gravitino} or the axino\cite{axino} as 
the LSP and the candidates for dark matter particle, stau 
can serve as the NLSP in the similar way. 
Our studies in this paper can be 
easily generalized into these cases.}. In this section we 
calculate the inclusive cross section of stau production from
the collision of high energy cosmic neutrinos on the earth matter.

The process which we are interested in is
\begin{equation}
\label{proc}
\nu+N\to \tilde{l}+\tilde{q}+X\to 2\tilde{\tau}+X'\ ,
\end{equation}
where $N=\frac{1}{2}(n+p)$ is an isoscalar nucleon, $X$ ($X'$) refers to 
any type of hadrons and leptons. 
At the parton level the corresponding process is 
$\nu+q\to \tilde{l}+\tilde{q}$
with the exchanges of the chargino, $\chi^+$, or neutralino, $\chi^0$
in the $t$-channel,
where $q$ stands for the valence and sea quarks.
The $\tilde{l}$ and $\tilde{q}$ in Eq. (\ref{proc}) 
will quickly decay into two NLSPs. 
For the calculation of the inclusive cross section in (\ref{proc}), 
we use the automatic Feynman Diagram Calculation
package(FDC) \cite{wang:fdc96}, which adopts the CTEQ6-DIS parton distribution
function\cite{cteq6} for the isoscalar nucleon. 


\begin{figure}
\includegraphics[scale=0.6]{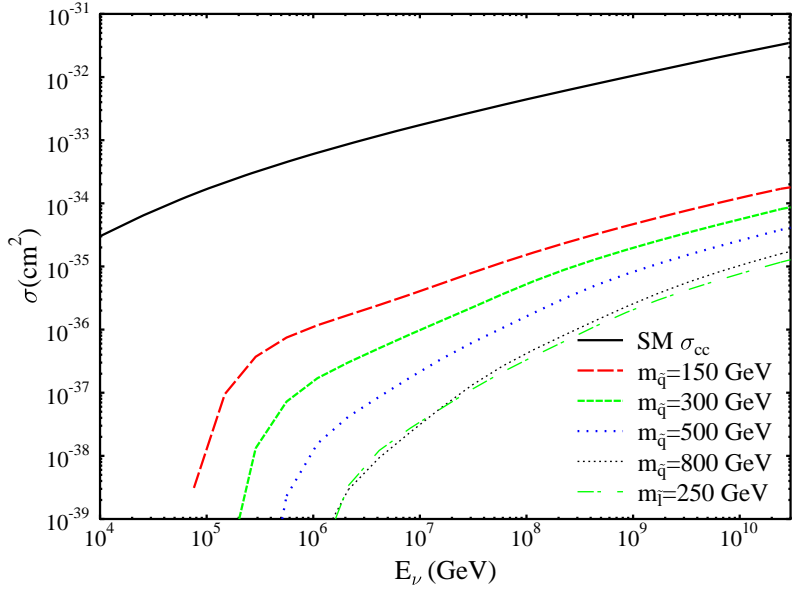}
\caption{\label{cross}
The cross sections of the inclusive process Eq. (\ref{proc}) for $M_1=M_2
=\mu=250 GeV$, $m_{\tilde{L}}=m_{\tilde{e_R}}=150 GeV$, and 
$m_{\tilde{Q}}=m_{\tilde{u_R}}=m_{\tilde{d_R}}=150, 300, 500, 800 GeV$.
The curve with the smallest cross section is for $M_1=M_2
=\mu=400 GeV$, $m_{\tilde{L}}=m_{\tilde{e_R}}=250 GeV$, and
$m_{\tilde{Q}}=m_{\tilde{u_R}}=m_{\tilde{d_R}}=800 GeV$.
The cross section for the standard model charge current process is also
plotted in the figure.
}
\end{figure}

For the soft SUSY breaking terms we take the parameters as 
$m_{\tilde{L}}=m_{\tilde{e}_R}=150 GeV$
in the calculations, which gives rise to a lighter stau of
about $143 GeV$. The gaugino masses and Higgsino
mixing parameter are taken as $M_1=M_2=\mu=250 GeV$, so that the neutralinos
and charginos are heavier than $\tilde{\tau}$. These parameters give
the lightest neutralino mass $m_{\chi^0}=181 GeV$ and the lighter
chargino mass $m_{\chi^+}=193 GeV$. In Fig. \ref{cross},
we plot the cross sections for $m_{\tilde{Q}_L}=m_{\tilde{q}_R}=150,
300, 500, 800 GeV$, which are the most sensitive supersymmetric parameters
to the cross section.
Another set of the soft parameters are taken as 
$m_{\tilde{L}}=m_{\tilde{e}_R}=250 GeV$, $M_1=M_2=\mu=400 GeV$ and
$m_{\tilde{Q}_L}=m_{\tilde{q}_R}=800 GeV$. (We will denote this set
of parameters by $m_{\tilde{l}}=250 GeV$ in the text.) 
In the figure, the inclusive cross section of the charged current 
process $\nu_\mu+N\to \mu+$ anything is also plotted. One can see that the
supersymmetric cross section is about 3 orders of magnitude smaller than 
the standard
model one.


\section{The distribution of NLSP energy and its range}

When the staus propagate in the earth they lose energy through ionization
and radiation processes: bremsstrahlung, pair production,
and photonuclear interactions. While the energy loss due to ionization 
is only slowly 
logarithmically increasing with the energy, the radiation processes cause 
a loss which in the high energy limit is proportional to the stau 
energy. The energy loss is usually expressed as
\begin{equation}
-\frac{dE_{\tilde{\tau}}}{dX}=\alpha+\beta E_{\tilde{\tau}} \ ,
\end{equation}
where $\alpha$ is due to ionization and $\beta$ is due to radiation.
We take $\alpha=2\times 10^{-3} GeV g^{-1} cm^2$,
which is the same as the ionization loss coefficient of $\mu$ 
in the rock\cite{pdg}.
Since $\beta$ is proportional to the inverse square of the incident
particle mass, we take $\beta=\beta_\mu\cdot
\left(\frac{m_\mu}{m_{\tilde{\tau}}}\right)^2$. 
For $\mu$, $\beta_\mu=3.9\times 10^{-6} g^{-1} cm^2$\cite{pdg}.
We then obtain the range of the stau with initial energy $E_0$
\begin{equation}
R(E_0)=\frac{1}{\beta}\log\frac{\alpha+\beta E_0}
{\alpha+\beta m_{\tilde{\tau}}}\ ,
\end{equation}
where we have assumed that the energy threshold of the detector can be very
small.

The range of the stau is actually a distribution as a function of its 
energy.
For simplicity, we use the range for the average energy of stau,
$R(\langle E_{\tilde{\tau}}\rangle)$, in the calculation of the stau flux.
 Since the collision is through $t$-channel, the stau
from the slepton decay and that from the squark decay have very different
energies. To discriminate them, we denote them by $\tilde{\tau}_1$ and 
$\tilde{\tau}_2$, and the corresponding rang $R_{\tilde{\tau}_1}$ and 
$R_{\tilde{\tau}_2}$ respectively.

\begin{figure}
\includegraphics[scale=0.6]{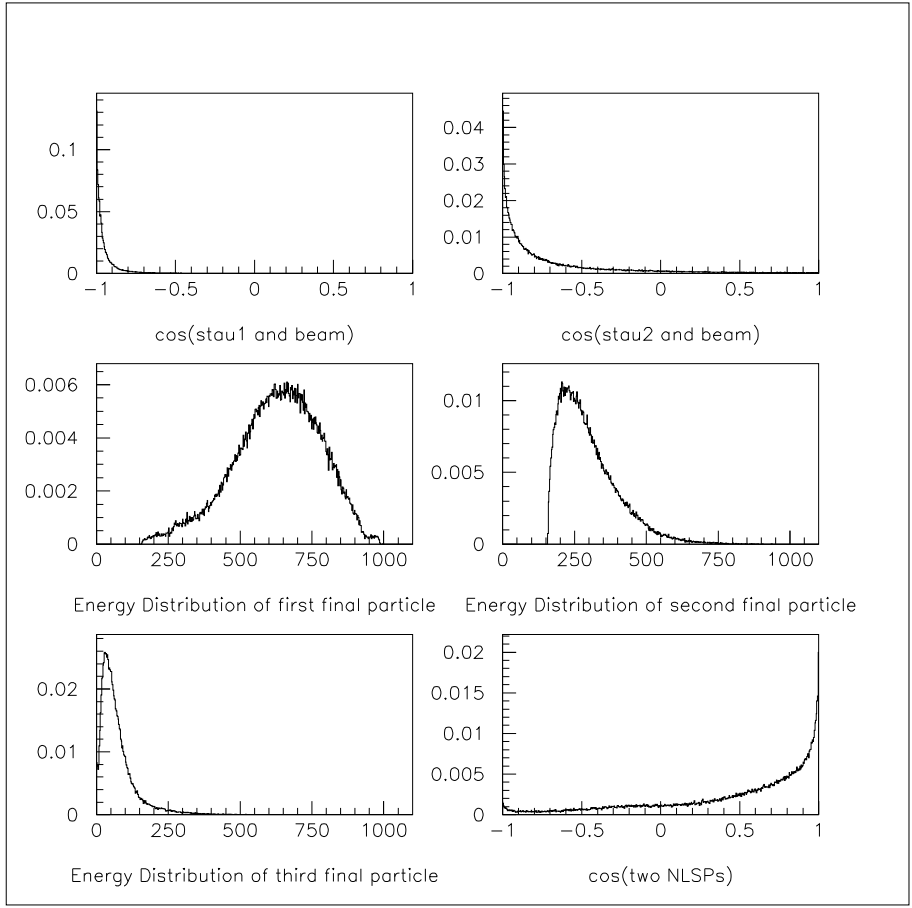}
\caption{\label{distri}
Angular ($\cos\theta$) and energy distribution of the final particles 
in the $\nu+N$ CM system, with
the beam direction referring to the direction of the incident nucleon.
The figures represent the angular distribution of $\tilde{\tau}_1$
and $\tilde{\tau}_2$ with the beam direction and the energy distributions
for $\tilde{\tau}_1$, $\tilde{\tau}_2$, $l'$ respectively and  the angle
between the two staus.
We take the CM energy of $\nu$ and $N$ at 1000 GeV. At higher energies 
the angular distribution of $\tilde{\tau}_{1,2}$ is more concentrated 
at $-1$ and the angle between them becomes smaller, while the energy
distribution is almost not changed.
}
\end{figure}

For $\tilde{\tau}_1$, we assume it has the same energy as the 
initial slepton, since all the
three flavor sleptons have the similar masses. While for $\tilde{\tau}_2$, 
we have to calculate the energy distribution
for the process $\tilde{q}\to \tilde{\tau}_2+l'+q'$.
In our calculation the energy distributions for the process
$\nu+N\to \tilde{\tau}_1+\tilde{\tau}_2+ q'+l' +X $ are obtained
by using the Monte
Carlo method to calculate the four-body final state. 
In Fig. \ref{distri} we plot the angular and the energy distribution of 
the final state particles of the process above in the $\nu+N$ CM system. 
The angle between the two stau is plotted in the last panel.
From the figure we see that the two stau are almost in the same 
direction as the incident neutrinos. 
This is easy to be understood since
the struck quark carries only small fraction of the total momentum of
the nucleon. As the energy increases, the $\tilde{\tau}_1$  and 
$\tilde{\tau}_2$ are more concentrated in the forward direction
of the incident neutrinos. Therefore, the transverse momentum of the 
slepton
in the CM system is generally much smaller than the momentum of $\nu$ and
$N$. Considering that the momentum of the struck quark is widely distributed
from just above the threshold to the total momentum carried by the nucleon,
we find it is difficult to estimate the transverse momentum of the slepton.

In the $\nu+N$ CM system, the average energy of $\tilde{\tau}_1$ 
changes from $\sim 65\%E_\nu^*$ to $\sim 75\%E_\nu^*$, while the average 
energy of $\tilde{\tau}_2$ changes from $\sim 15\%E_\nu^*$ 
to $\sim 35\%E_\nu^*$, both depending not sensitively on the energy of the
neutrino,
$E_\nu^*$, and the squark mass. We fix $E_{\tilde{\tau}_1}^*=70\%E_\nu^*$,
$E_{\tilde{\tau}_1}^*=20\%E_\nu^*$ in the CM systerm as a simplification
of the calculation. Assuming that 
the two stau are both in the forward direction, we boost them to the 
laboratory system. We then get 
\begin{equation}
E_{\tilde{\tau}_{1(2)}}^{\text{lab}}\approx 
\frac{E_\nu^*}{m_N} (E_{\tilde{\tau}}^*+|p_{\tilde{\tau}}^*|)
\approx \frac{2E_\nu^*}{m_N} E_{\tilde{\tau}}^* \approx
70\% (20\%) E_\nu\ ,
\end{equation}
where $E_\nu$ is the energy of the primary cosmic neutrinos,
and we ignore the $\tilde{\tau}$ mass in the second step, which is
a rough approximation at low energies.

\section{ event rate}
\subsection{Neutrino flux}

Waxman and Bahcall (WB) have set an upper bound on the high energy 
diffuse neutrino flux from the observed cosmic ray flux at high energies.
Depending on the evolution of the source activity, they gave the limit
on muon neutrino and anti-neutrino extra-galactic flux\cite{wb}
\begin{equation}
\label{wb1}
\left(  \frac{d\phi_\nu}{dE} \right)_{\text{WB}}
=(1\sim 4)\times 10^{-8} \cdot\left( \frac{1 GeV}{E}\right)^2
GeV^{-1} cm^{-2} s^{-1} sr^{-1}\ .
\end{equation}
If ``unknown source'' of protons is taken into account, the low energy
neutrino flux can be raised as $d\phi_\nu/dE\propto E^{-2}
[1+0.1((10^8 GeV)/E)]$\cite{wb}. However, the neutrino flux at the energies 
below about $\sim 10^6 GeV$
is bounded by AMANDA experiment\cite{amanda}, i.e., $d\phi_\nu/dE
\le 8.4\times 10^{-7}E^{-2} GeV cm^{-2} s^{-1} sr^{-1}$. 
We refer to the maximal value of flux
(\ref{wb1}) as `WB1' flux, while the flux raised at low energies 
and bounded by AMANDA experiment below $\sim 10^6 GeV$ as the `WB2' flux.

The WB bound holds for the optically thin neutrino sources. Mannheim, 
Protheroe and Rache (MPR) extend the calculation to ``optically thick''
sources to the nucleons\cite{mpr}. They give an upper bound of diffuse 
neutrino flux which
is almost 30 times larger than the WB1 limit in Eq. (\ref{wb1}).

When the extra-galactic neutrinos propagate to the earth, we expect they
contain the three flavor neutrinos in ratio of 1:1:1 
due to the neutrino oscillation. The supersymmetric particle production
rate should be independent of the neutrino flavor since we take all the
sleptons almost degenerate. Therefore the three flavors of neutrinos should
contribute with the same signal rate, except that $\bar{\nu}_e$ has more strongly
scattering effect with the earth electron. Here, we calculate the supersymmetric
particle signals at the detector contributed by the $\nu_\mu+\bar{\nu}_\mu$
flux with WB1, WB2 and MPR limit.

\subsection{Calculation}

\begin{figure}
\includegraphics[scale=0.6]{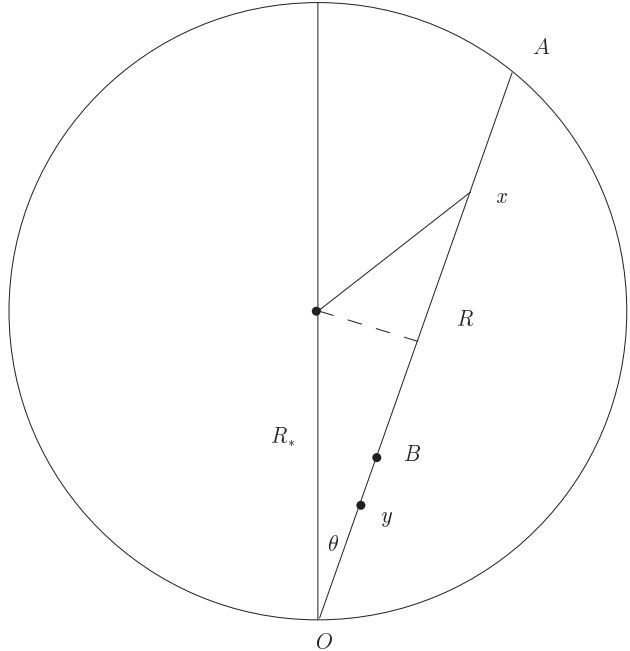}
\caption{\label{earth}
The path of neutrino flux penetrates the earth 
to the detector at the angle $\theta$.
}
\end{figure}

The earth becomes opaque for neutrinos above about $40 TeV$ from
the nadir due to the charged-current interaction between neutrino and
nucleon. Therefore the scattering of $\nu_\mu$ to $\mu$ has to be taken
into account.

In Fig. \ref{earth} we show the path of a neutrino flux penetrating the
earth before it reaches the detector located at the point $O$. 
We assume that the
primary flux at the point $A$ is $\phi_{\nu_0}$, then the flux at the
point $x$ is given by
\begin{equation}
\phi_{\nu}(x)=\phi_{\nu_0} e^{-\int_A^x n(r)\sigma_\mu dR }\ ,
\end{equation}
where the integration is along the path from the point $A$ to $x$,
with $n(r)$ the nucleon number density,
$\sigma_\mu$ the standard model charged-current cross section
and $R$ is between $0$ and $2R_*\cos\theta$, with
$R_*=6371$ km being the radius of the earth.
The number density $n(r)$ is given by $n(r)=\rho(r)N_A$, 
where $\rho(r)$ is the density
profile of the earth given in Ref. \cite{gandhi} and $N_A$ is the
Avogadro's number. 

Assuming the stau produced below the point $B$ can reach the detector, 
we then get the flux of stau at the detector $O$ 
\begin{eqnarray}
\label{tauflux}
\phi_{\tilde{\tau}}
&=& \int_B^O \phi_\nu(B)
e^{-\int_B^y n(r)\sigma_\mu dR }
\sigma_{\tilde{\tau}} n(r)dy\nonumber\\
&=&\phi_\nu(B) \frac{\sigma_{\tilde{\tau}}}{\sigma_\mu}
\left( 1-e^{-N_A\sigma_\mu R_{\tilde{\tau}} }\right)\ ,
\end{eqnarray}
where both integration are along the path and $\sigma_{\tilde{\tau}}$ is
the supersymmetric cross section.
It should be noted that at the point corresponding to $R_{\tilde{\tau}_1}$
there is only one stau finally arriving at the point $O$, while at the
point corresponding to $R_{\tilde{\tau}_2}$, there can be two
stau arriving at the point $O$.

At large angles, the range of the stau may be larger than the actual
depth of the earth that a neutrino has penetrated. 
In this case we take the actual depth the neutrino
penetrated to calculate
the probability that a stau can be produced by a neutrino.

\begin{figure}
\includegraphics[scale=0.6]{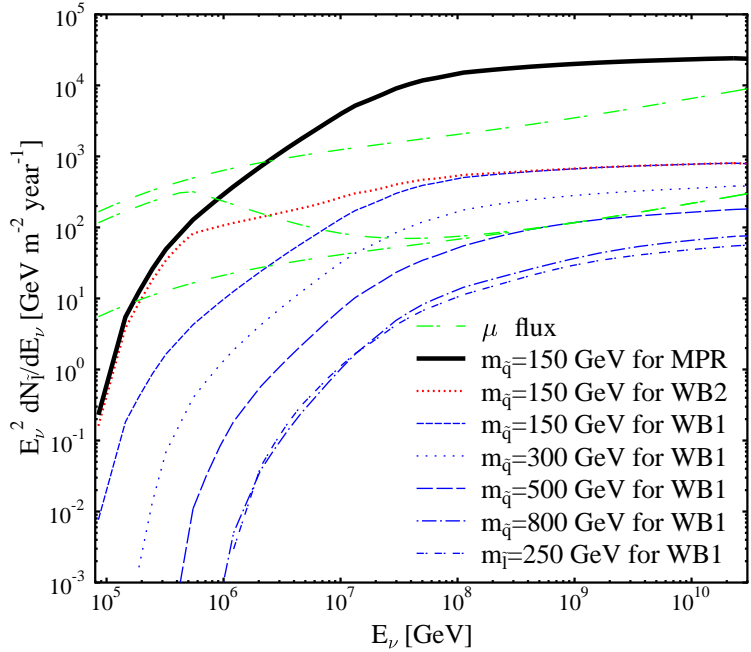}
\caption{\label{flux}
Flux $E_\nu^2 {d\phi_{\tilde{l}}}/{dE_\nu}$ 
as a function of the incident neutrino energy for WB1, WB2 and MPR 
neutrino fluxes and $m_{\tilde{q}}=150, 300, 500, 800 GeV$.
The flux for the set of soft parameters with $m_{\tilde{l}}=250 GeV$
is also plotted.
Similar quantity for $\mu$ flux is plotted in figure. The three
curves for $\mu$ flux calculated for the MPR, WB2 and WB1 neutrino fluxes
 respectively are shown from the top to the bottom.
}
\end{figure}

In Fig. \ref{flux}, we plot the flux $E_\nu^2 {d\phi_{\tilde{l}}}/{dE_\nu}$ 
as a function of the incident neutrino energy for WB1, WB2 and MPR 
neutrino fluxes and for $m_{\tilde{q}}=150, 300, 500, 800 GeV$.
For $m_{\tilde{q}}=150 GeV$
we show the results of WB1, WB2 and MPR fluxes.
For other parameters we plot only the result of WB1 in the figure for 
clarity. 
We notice that the curves become almost flat at the energy above
$\sim 10^8 GeV$. The reason is that for the energies above
$\sim 10^8 GeV$ the range of $\tilde{\tau}$ is so large that
it is comparable with the radius of the earth. Therefore in most 
directions the $\phi_{\tilde{\tau}}$ is proportional to the primary neutrino
flux according to Eq. (\ref{tauflux}), i.e., $\phi_\nu(B)=
\phi_{\nu_0}$ for $R_{\tilde{\tau}} > R(\theta)$.

Note that the contribution to the total $\tilde{\tau}$ events comes mainly
from the low energy neutrino flux. (That will be very clear in a 
$dN_{\tilde{l}}/dE_\nu - E_\nu$ figure.)
The reason is that the neutrino flux decreases with energy with the power law
index of $-2$, while the supersymmetric cross section increases with energy
(well above the threshold)
with a smaller power law index of only about $0.4$.

The corresponding curves for the $\mu$ flux is also plotted in the
figure for the WB1, WB2, and MPR neutrino fluxes.
It is quite noticeable that the stau flux is even comparable with the
$\mu$ flux in the case of $m_{\tilde{q}}=150 GeV$. 

\begin{table}
\begin{tabular}{|l||c|c|c|c|c|c|}
\hline
& $\mu$ & ($m_{\tilde{q}}=$)  $150 GeV$ & $300 GeV$ & $500 GeV$ & $800 GeV$ & $m_{\tilde{l}}= 250 GeV$ \\
\hline
WB1 & $1.8\times 10^{-4}$ &$6.6\times 10^{-5}$  & $1.5\times 10^{-5}$ & $3.3\times 10^{-6}$  & $6.5\times 10^{-7}$ & $5.8 \times 10^{-7}$\\
\hline
WB2 & $3.2\times 10^{-3}$ &$3.7\times 10^{-4}$ & $4.9\times 10^{-5}$ & $7.0\times 10^{-6}$ & $1.0\times 10^{-6}$ & $9.4\times 10^{-7}$ \\
\hline
MPR& $5.4 \times 10^{-3}$ &$2.\times 10^{-3}$ & $4.4\times 10^{-4}$ & $1.0\times 10^{-4}$ & $1.9\times 10^{-5}$ & $1.7\times 10^{-5}$ \\
\hline
\end{tabular}
\caption{
\label{result}
Event rate of stau at the detector for per year per $m^2$.}
\end{table}

Finally, the total event rate for one year is given by
\begin{equation}
\text{event rate}=1\text{year}\cdot 2\pi\int \phi_{\tilde{\tau}}
dE_\nu d\cos\theta \ .
\end{equation}
In table \ref{result}, we give the event rates per year per $m^2$ for
the WB1, WB2 and MPR fluxes for different squark masses.
The total $\mu$ events per year per $m^2$ is also given in the table.
It is quite interesting to notice that even in the most 
conservative case we may observe one such a event
at a detector with  $1 km^2$ effective area per year.

\section{conclusion}

In this paper we have considered the phenomenology of the 
model with quintessino as the lightest supersymmetric
particle and the dark matter candidate\cite{quintessino}. 
In this scenario
there exists a long-lived charged heavy particle, usually the lighter
$\tilde{\tau}$. It is quite possible to detect such particles in cosmic
ray detectors with large effective area. Discovery of the
supersymmetric particles of cosmic source will be a valuable 
complementary for supersymmetry search at high energy colliders. 



\begin{acknowledgments}
We thank G. Burdman, Shuwang Cui, Linkai Ding, J. Feng, Tao Han and 
Yi Jiang for discussions.  This work is
supported in part by the National Natural Science Foundation of China
under the grant No. 10105004, 19925523, 90303004
and also by the Ministry of Science and Technology of China under
grant No. NKBRSF G19990754. X. J. Bi is also supported
in part by the China Postdoctoral Science Foundation.
\end{acknowledgments}

\end{document}